\documentclass[]{spie}  

\usepackage{amsmath,amsfonts,amssymb}
\usepackage{subcaption}
\usepackage{graphicx}
\usepackage[colorlinks=true, allcolors=blue]{hyperref}
\usepackage{rotating}

\title{Machine Learning Based Alignment For LCLS-II-HE Optics}

\author{Aashwin Mishra}
\author{Nicholas Brennan}
\author{Tianyu Huang}
\author{Jason Jaquith}
\author{Hasan Yava\c{s}}
\author{Matthew Seaberg}
\affil{SLAC National Accelerator Laboratory, Menlo Park, CA 94025, USA}

\authorinfo{Further author information: (Send correspondence to Matthew Seaberg)\\Matthew Seaberg: E-mail: seaberg@slac.stanford.edu}

\begin{document} 
\maketitle

\begin{abstract}
The hard X-ray instruments at the Linac Coherent Light Source are in the design phase for upgrades that will take full advantage of the high repetition rates that will become available with LCLS-II-HE. The current X-ray Correlation Spectroscopy instrument will be converted to the Dynamic X-ray Scattering instrument, and will feature a meV-scale high-resolution monochromator at its front end with unprecedented coherent flux. With the new capability come many engineering and design challenges, not least of which is the sensitivity to long-term drift of the optics. With this in mind, we have estimated the system tolerance to angular drift and vibration for all the relevant optics ($\sim$10 components) in terms of how the central energy out of the monochromator will be affected to inform the mechanical design. Additionally, we have started planning for methods to correct for such drifts using available (both invasive and non-invasive) X-ray beam diagnostics. In simulations, we have demonstrated the ability of trained Machine Learning models to correct misalignments to maintain the desired central energy and optical axis within the necessary tolerances. Additionally, we exhibit the use of Bayesian Optimization to minimize the impact of thermal deformations of crystals as well as beam alignment from scratch. The initial results are very promising and efforts to further extend this work are ongoing.

\end{abstract}


\section{INTRODUCTION}
\label{sec:intro}  

The Linac Coherent Light Source (LCLS), the first hard X-ray free electron laser in the world, has recently completed the LCLS-II upgrade and is ready to be turned on to deliver high repetition rate at soft X-ray energies. The LCLS-II-HE project is currently in the design phase and will extend LCLS-II to hard X-ray energies, eventually reaching 20 keV. With LCLS-II-HE, we have an opportunity to make measurements that are orders of magnitude more sensitive than what can be made currently at LCLS. This can take many shapes and forms, including high-resolution spectroscopic measurements at the upcoming Dynamic X-ray Scattering (DXS) instrument, all the way to imaging operando systems with high resolution by taking advantage of the unprecedented coherent flux\cite{Schoenlein2016}. 

All of these new directions rely on sensitive (and sometimes complex) optical systems that are often spread over many meters and must be aligned properly and efficiently. Furthermore, in order to ensure successful experiments the proper alignment must be maintained within a tight tolerance over the course of hours to days. These are tasks that are increasingly challenging for a human expert to perform without help. Here, we present first steps towards the use of Machine Learning (ML) methods for the alignment and drift correction of a model x-ray optical system. The system chosen for the study is a novel hard X-ray high-resolution monochromator (HRM), designed to preserve the bandwidth-limited XFEL pulse duration. The system is based on the zero dispersion stretcher concept developed for ultrafast lasers, the details of which will be the subject of a forthcoming article and not described here. As the system currently only exists at the conceptual level, all of the results presented in the following sections are based on wave optics simulations rather than physical experiments.

For the purposes of this manuscript, this monochromator design provides a case study of a system with many degrees of freedom and X-ray beam diagnostics, with a photon energy output that is strongly sensitive (relative to the desired $\sim$meV resolution) to misalignments of a number of the relevant degrees of freedom. Furthermore, with the LCLS-II-HE upgrade, the anticipated increase in repetition rate is expected to eventually produce $>$100W of average power, meaning that in some cases individual optics (crystals, mirrors, etc) can absorb 10's of Watts of beam power. This will result in thermal deformation of the optical surface that may have a sizable impact on the XFEL wavefront, thus impacting the beam focusing quality which in turn will degrade the monochromator resolution\cite{Zhang2023}. All of these effects can be simulated, the results of which can be used to influence the design of the system, as well as to inform and test mitigation strategies such as those described here.

In this study, we considered three distinct test cases. First, we used Bayesian Optimization to mitigate the simulated impact of crystal thermal deformation under a high heat-load. Second, we exhibit the potential of Bayesian Optimization for the alignment of a section of the monochromator optical system from scratch. Finally, we investigated the use of a Support Vector Machine in a feedback loop to compensate for simulated optical misalignments of the monochromator consistent with expected thermal drift on the minutes to hours timescale, including 5 misalignment degrees of freedom as well as 6 diagnostic devices that are included in the design.

The manuscript is arranged as follows: after a detailed overview in Section I, in Section II we outline some of the concepts from data-driven modeling and control that are utilized in this investigation. Then, in Section III, we outline each of our test cases, report the results therefrom, and discuss the same. The manuscript concludes with a summary and outline of future steps in Section IV.

\section{Mathematical Details}
In this section, we provide an overview of the concepts and techniques utilized in this investigation. The objective is to ensure that domain experts are introduced to the Machine Learning approaches utilized in this investigation, so as to ensure that this communication is self-contained. Interested readers are referred to other literature\cite{hastie2009elements, murphy2012machine, murphy2022probabilistic} for detailed discussions.

\textit{Machine Learning} (ML) is a branch of artificial intelligence that specifically focuses on enabling computer-based systems to infer predictive models from data. The notion of learning expresses the ability of such an algorithm to progressively improve its performance in a specific task by processing relevant data and information. Traditionally, algorithms do not rely on explicit rules from domain experts for how to perform the task. However, a growing branch of ML demonstrates that learning can be improved by merging ML algorithms with domain expertise. Depending on the context, this has been referred to as Domain Aware Machine Learning or Physics Informed Machine Learning. In ML, supervised learning encompasses methods that learn from a collection of labeled data, $\{(x,y)_i\}^{N_l}_{i=1}$ to predict an outcome $\hat{y}$ for an input $\hat{x}$, through application of a learning hypothesis $f(\hat{x})$. Common applications of supervised learning include classification and regression problems.

\textit{Bayesian Optimization}\cite{frazier2018bayesian} (BO) is a sequential sampling approach for finding global optima of black-box functions that are expensive to sample from. It uses a surrogate probabilistic model and the location of every subsequent sample in the sequence is determined using an acquisition function over the probabilistic prediction of this surrogate model. The key advantage of Bayesian Optimization is that it allows optimization in minimal evaluations of the underlying process. This is useful both for time-consuming simulations as well as precious LCLS beamtime. Secondly, it enables us to handle general black box functions without assuming any functional form, via the use of flexible Bayesian models like Gaussian Processes and Bayesian Neural Networks. This is advantageous as we do not have analytical descriptors of the LCLS-II-HE optical system alignments. Finally, owing to the use of probabilistic surrogate models, Bayesian Optimization is robust to noise in the function evaluations (e.g. the noise inherent to minimally-invasive diagnostics). In this vein, Bayesian Optimization has been used for many complex applications such as optimizing the architecture and hyperparameters of neural networks.

A \textit{Gaussian Process}\cite{rasmussen2006gaussian} (GP) is a non-parametric model that calculates probability densities over the space of functions, providing a pragmatic and probabilistic approach to kernel machine learning. While a Gaussian distribution is characterized by a normal distribution with mean and covariance($Y \sim \mathcal{N}(\mu,\,\sigma^{2})$), a Gaussian process is defined via a normal distribution over mean and covariance functions, $Y \sim GP(m(X),k(X,X'))$. Herein, $m(X)$ and $k(X,X')$ are referred to as the mean and the covariance functions. GP models are attractive in many fields as they can represent various data complex structures. Due to these advantages, they are regarded as the default proxy model in Bayesian Optimization, as well as other Active Learning studies.

A \textit{Support Vector Machine} (SVM) is  a non-probabilistic algorithm that forms decision boundaries within linearly separable data. In a binary classification problem, these decision boundaries are formed by solving for the maximum distance between points closest to the decision boundary. These points are called support vectors. SVMs employ hyperplanes for classifying and regressing data. SVMs can be extended to nonlinear problems by transforming the data points to a higher-dimensional space, rendering the data linearly separable. This transformation is usually accomplished with the use of inner products. However, since the computational costs can become intractable with large datasets, the inner product is approximated using a kernel function in practical algorithms, a process known as kernelizing. The radial basis function kernel, constructed based on the Euclidean distance of vectors, is a popular kernel for nonlinear data.

\textit{Deep Learning}\cite{bishop2006pattern} is a class of ML in which multilayer representations are utilized to extract hierarchical features from a complex input. Common deep learning models are based on neural network architectures and include densely connected or fully connected neural networks, Convolutional Neural Networks (CNNs), Recurrent Neural Networks (RNNs), and deep belief networks. Adhering to the fully connected neural network as an archetype, the neurons of adjacent layers are densely connected, and outputs of a preceding layer are fed forward as inputs to the succeeding layer. In this vein, a fully connected neural network defines a mapping from the space of the input layer, $z_0$, to the output layer, $z_l$. The layers between the input and output are referred to as hidden layers by convention. Mathematically, any two adjacent layers in such a network are connected via $z_l=g_l(W^T_lz_{l-1}+b_l)$, where $g_l$ denotes the activation function associated with the layer, for which a number of options can be chosen, for instance, sigmoid, Rectified Linear Units (ReLU), hyperbolic tangent, etc. $W_l$ and $b_l$ are the weights and biases associated with the layer corresponding to the subscript $l$.

\section{Results \& Discussion}


   \begin{sidewaysfigure} 
   \begin{center}
   \begin{tabular}{c} 
   \begin{subfigure}{0.45\textwidth}
    \includegraphics[height=11cm]{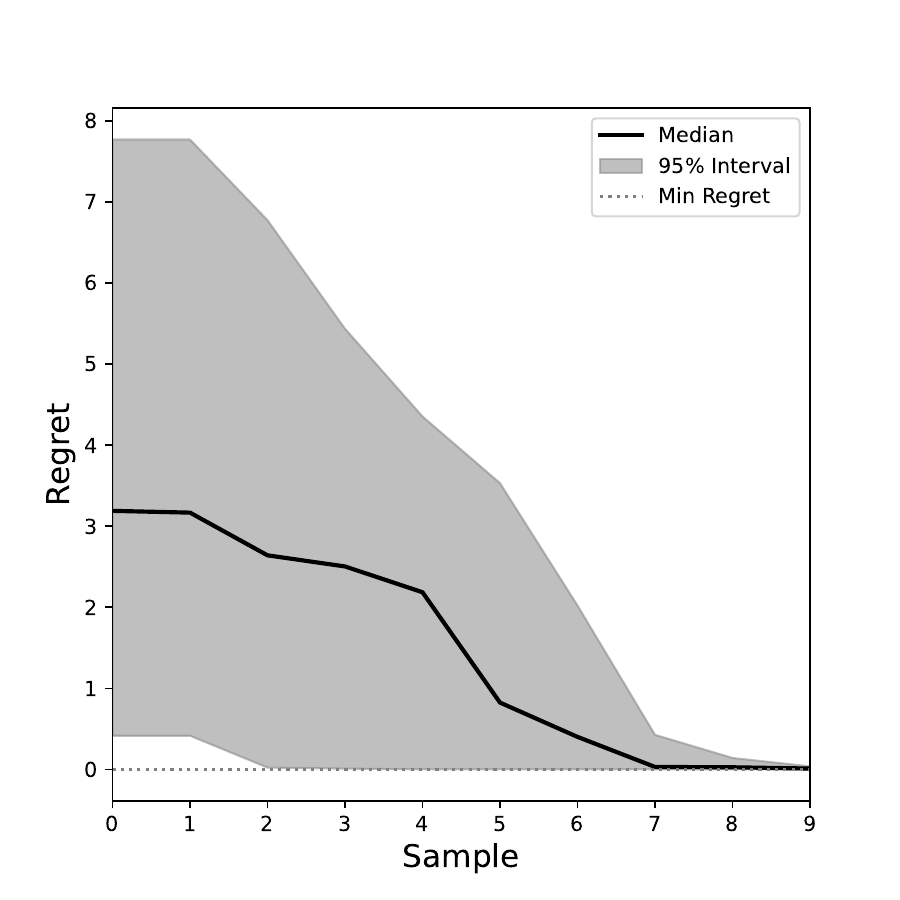}
    \caption{Bayesian Optimization} \label{fig:1a}
  \end{subfigure}%
  \hspace*{\fill} 
   \begin{subfigure}{0.45\textwidth}
    \includegraphics[height=11cm]{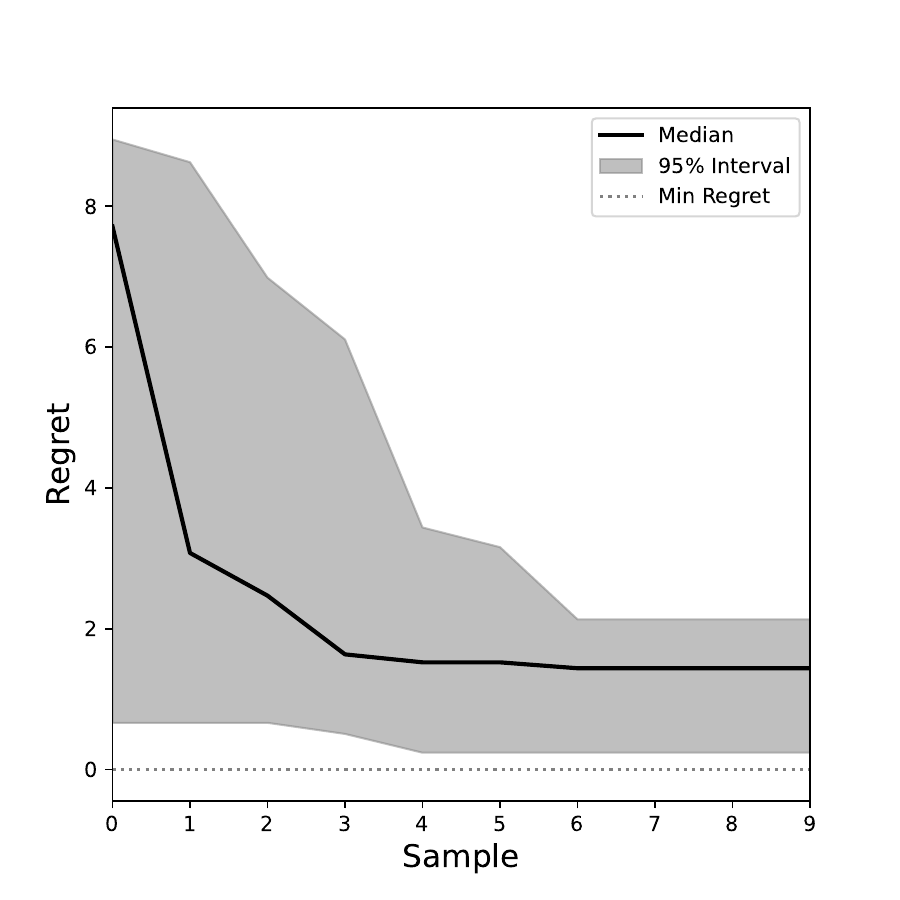}
    \caption{Random Sampling} \label{fig:1a}
  \end{subfigure}%
   \end{tabular}
   \end{center}
   \caption[example] 
   { \label{fig:example} 
Use of Bayesian Optimization to compensate for crystal thermal deformation in the DXS HHLM while using a bendable mirror: (a) The figure exhibits the median and the $95\%$ interval over a sequence of 100 Bayesian Optimization chains, (b) The figure exhibits the statistics over 100 chains of random sampling.}
   \end{sidewaysfigure} 

We have utilized Bayesian Optimization to simulate minimization of the impact of thermal deformations on a representative crystal in the HRM, informed by thermal finite element analysis (FEA) simulations. Figure 1 shows an example optimization of the beam wavefront through the monochromator. For this test case, the FEA was performed assuming 40 W power absorption within the beam footprint by a silicon crystal cooled to the coefficient of thermal expansion crossover point near 125 K\cite{Zhang2023}. The wavefront simulation involved reflection from the Si 111 lattice plane of the deformed crystal, while allowing for bending adjustment of the collimating mirror upstream of the monochromator to compensate for the spherical component of the thermal deformation. The overall impact on the XFEL wavefront was evaluated by analyzing the focusing performance upon reflection from a perfect focusing mirror. For the Bayesian Optimization, we utilized a Gaussian Process Regressor as our probabilistic proxy model with a  Matern $3/2$ Kernel and a constant mean. For the acquisition function, we utilize Expected Improvement. As can be seen in Figure 1, Bayesian Optimization outperforms random sampling with respect to the absolute value of the optimum and the number of samples required to achieve it. Here, $regret_T(\mathcal{A}) = \sup_{\{f_1,...,f_T\}} \sum_{t=1}^{T}f_t(x_t) - \min_{x in \mathcal{K}} \sum_{t=1}^{T} f_t (x)$. In practice, we envision this optimization will be performed using a wavefront sensor to optimize the collimation.

   \begin{figure} [ht]
   \begin{center}
   \begin{tabular}{c} 
    \includegraphics[width=0.75\textwidth]{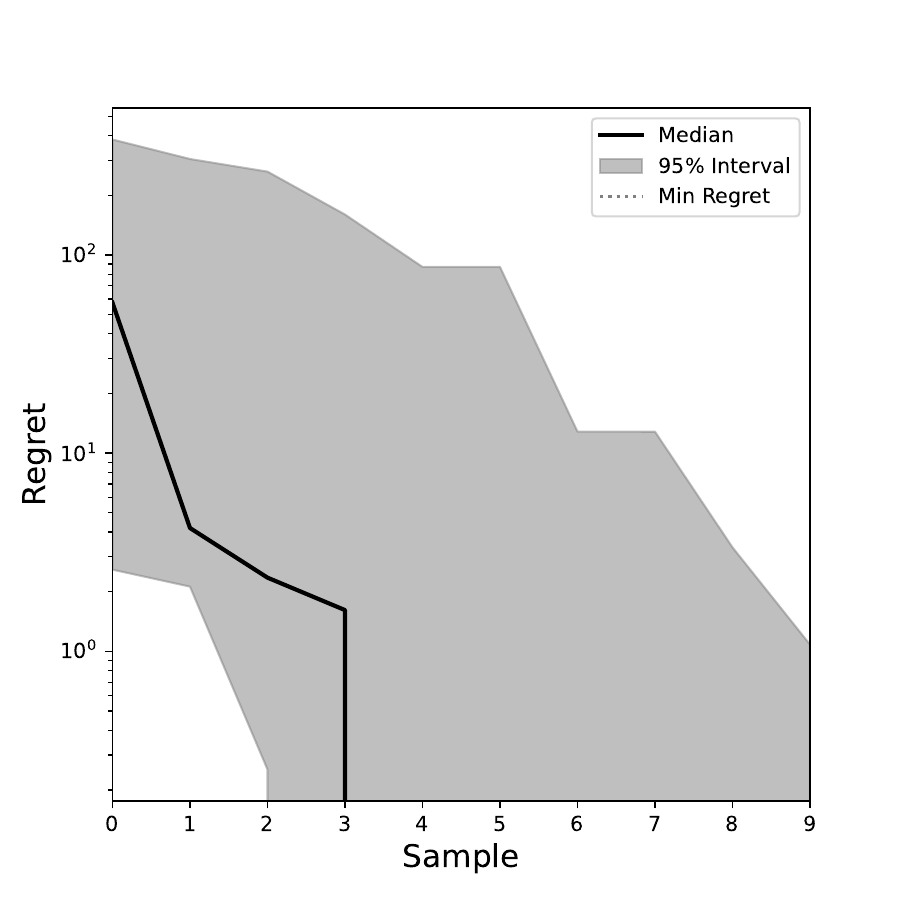}
   \end{tabular}
   \end{center}
   \caption[example] 
   { \label{fig:example} 
Bayesian Optimization for alignment of a section of the monochromator optical system from scratch. The figure exhibits the median and the $95\%$ interval over a sequence of 50 Bayesian Optimization chains.}
   \end{figure} 

Our second test case deals with correcting the beamline from a misaligned state, where devices were misaligned by substantial changes to the glancing angle of incidence from the optimal alignment. There were four inputs to the beamline that were misaligned, explicitly the angle and focal lengths of two mirrors used to expand and collimate the beam horizontally using a Wolter Type II geometry, which condition the beam's spatial properties for input to the HRM. The objective was to maximize the summed intensity on two simulated imagers located downstream of the mirrors. In the Bayesian Optimization, we utilized a Gaussian Process Regressor as our model with a Matern $3/2$ Kernel and a constant mean. For the acquisition function, we utilize Expected Improvement. The results over $50$ BO chains are outlined in Figure 2. As can be seen, within a few samples, almost all the BO chains reach the optimal state.

Due to its stringent resolution requirement, HRM is susceptible to environment-related drift over time. With alignment tolerances down to the order of 10nrad, robust solutions are required for quickly diagnosing beam misalignment and prescribing realignment
procedures to counteract drifts in the beam path. We have tested and evaluated the efficacy of Machine Learning models to detect and correct for deviations from the optimal state of the mirrors along the HRM. These models subscribed to the Classification And Regression Trees (CARTs), Random Forests, Support Vector Machines (SVM) and Neural Network algorithms. The models were trained and tested using simulations, where the inputs were simulated measurements from five position-sensitive detectors and a spectrometer, while the model’s outputs were corrected settings for five mirrors. These trained models were further tested for their efficacy in realigning the simulated beam over iterations of such corrections. In this study, we found that ML models stabilize and realign the beam, and optimize the beam’s central energy within a few iterations as illustrated in Figure 3. In our preliminary tests, we found the Support Vector Machine based model to perform well and the realignment results correspond to this model.  

   \begin{sidewaysfigure}
   \begin{center}
   \begin{tabular}{c} 
    \includegraphics[width=\textwidth, trim={6cm 5.5cm 6cm 5.5cm},clip]{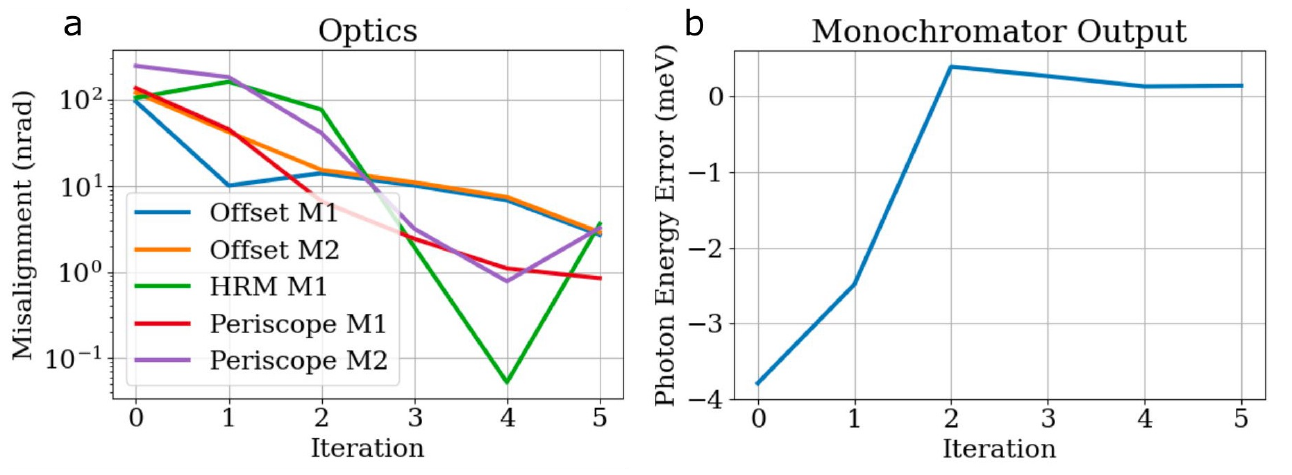}
   \end{tabular}
   \end{center}
   \caption[example] 
   { \label{fig:example} 
(a) A simulation was initialized with known misalignments of various optics within the system. Using a trained Support Vector Machine (SVM) regression model and the output of simulated beamline diagnostics, the system was brought back within alignment tolerances within 3 iterations. The resulting effect on the central energy of the monochromator output is shown in (b). Note that at the point the optics are back within tolerance, the error in photon energy is below the goal of 1meV.}
   \end{sidewaysfigure}

\section{Summary \& Future Work}
This investigation focuses on the application of Machine Learning based models for beam alignment and misalignment correction. We have demonstrated the ability of Machine Learning models to correct misalignments to maintain the desired central energy and optical axis within the necessary tolerances, in simulations. Further, we exhibited the use of Bayesian Optimization to minimize the impact of thermal deformations of crystals as well as beam alignment from scratch. The initial results are very promising and efforts to further extend this work are ongoing. 
In future work, these efforts will be augmented with probabilistic ML models to ensure safety and robustness, ML Interpretability approaches to assess and optimize sensor placement and selection, and also Verification and Validation of these models.

\acknowledgments 
Authors thank Lin Zhang and Hengzi Wang for providing FEA simulations of crystals under thermal load, and Oliver Hoidn for useful discussions. Aashwin Mishra was supported by the SLAC ML Initiative. This work was supported in part by the U.S. Department of Energy, Office of Science, Office of Workforce Development for Teachers and Scientists (WDTS) under the Science Undergraduate Laboratory Internships Program (SULI). Tianyu Huang was supported by the LCLS Summer Internship Program. Use of the Linac Coherent Light Source (LCLS), SLAC National Accelerator Laboratory, is supported by the U.S. Department of Energy, Office of Science, Office of Basic Energy Sciences under Contract No. DE-AC02-76SF00515.

\bibliography{report} 
\bibliographystyle{spiebib} 

\end{document}